# The micro-foundations of an open economy money demand:

# An application to the Central and Eastern European countries

Claudiu Tiberiu ALBULESCU,[1] Dominique PÉPIN,[2] and Stephen M. MILLER[3*]

[1] Management Department, Politehnica University of Timisoara; [2] CRIEF, University of Poitiers;

[3] Department of Economics, University of Nevada, Las Vegas

**Abstract**

This paper investigates and compares currency substitution between the currencies of Central and Eastern European (CEE) countries and the euro. In addition, we develop a model with microeconomic foundations, which identifies difference between currency substitution and money demand sensitivity to exchange rate variations. More precisely, we posit that currency substitution relates to money demand sensitivity to the interest rate spread between the CEE countries and the euro area. Moreover, we show how the exchange rate affects money demand, even absent a currency substitution effect. This model applies to any country where an international currency offers liquidity services to domestic agents. The model generates empirical tests of long-run money demand using two complementary cointegrating equations. The opportunity cost of holding the money and the scale variable, either household consumption or output, explain the long-run money demand in CEE countries.

---

[*] Stephen M. MILLER, Department of Economics, University of Nevada, Las Vegas. E-mail: stephen.miller@unlv.edu.

# 1. Introduction

The standard open-economy money demand model uses a two-country portfolio balance model (e.g., Leventakis 1993). This macro-model does not include microeconomic foundations and, thus, is subject to the Lucas's (1976) critique.[1] Because of its static nature, the estimated money demand may appear unstable for modified monetary policy strategies.

This paper investigates and compares currency substitution between the currencies of CEE countries and the euro. As CEE countries move toward more financial integration with the European Union, standard theory suggests that CEE households should use an increasing share of euro money relative to their own domestic money. Two policy implications emerge from our study. The monetary authorities in the CEE countries should consider not only the opportunity cost of holding money, but also the effect of the exchange rate, which occurs even absent strong currency substitution. The CEE countries need the political will to join the euro zone, even though efforts must continue toward higher monetary integration.

The empirical studies of the money demand typically do not provide a micro-founded theoretical model to justify the specification of their empirical money demand functions.[2] In addition, these models test for currency substitution through money demand's sensitivity to the exchange rate. Such a framework imposes important limitations, since one cannot examine one phenomenon (currency substitution) independently of the other (exchange rate sensitivity). Moreover, these models examine open-economy money demand sensitivity without differentiating between currency substitution and currency complementarity.

We contribute to the existing literature in two ways. First, we develop a micro-founded model that describes a mechanism through which the exchange rate affects money demand, even absent currency substitution. That is, money demand responds to exchange rate fluctuations, even after removing currency substitution, because the exchange rate affects the liquidity service associated with foreign money holding. Indeed, our model measures currency substitution intensity without explicitly considering the exchange rate. In addition, the model captures both currency substitution and currency complementarity hypotheses, which enables the assessment of the currency substitution intensity.[3] Further, to capture recent economic circumstances, where interest rates went negative, we consider an additional opportunity cost of holding money (domestic or foreign), which keeps the overall opportunity cost positive, even where the interest rate itself becomes negative.

---

[1] According to Hueng (1998), Dreger et al. (2007) and Hsieh and Hsing (2009), the overall effect of the exchange rate on the domestic money demand is not straightforward. Moreover, it is not clear whether the level or the (expected) exchange rate variation should enter the money demand equation. In addition, different measures of variables that enter the money demand equation appear in empirical studies without explicit theoretical support (Hueng, 2000).
[2] Chen (1973), Miles (1978), Bordo and Choudri (1982) and Hueng (1998, 2000) are exceptions.

Our micro-founded model, which integrates the liquidity production function and a Constant Elasticity of Substitution (CES) consumption function, produces a money demand function close to Miles (1978), who, however, does not consider the consumption choice. Thus, we demonstrate that the Bordo-Choudri (1982) criticism of Miles's (1978) money demand function is not relevant, since the function does not reflect the omission of consumption or income.

Second, we parameterize our model to test the long-run sensitivity of an open-economy money demand. In particular, our model fits the CEE case, where the euro offers liquidity services to domestic agents.[4] After their transition from a centralized to a market-based economic system, the CEE countries joined the European Union (EU). Several CEE countries already belong to the euro area, while others continue the integration process. Investigating money demand in euro area candidate countries offers information about their degree of monetary integration and about the liquidity services provided by the domestic currency compared to the euro.

---

[3] Currency substitution is defined as "the tendency of residents to replace domestic money with foreign currencies in response to changes in their relative rate of return" (Filosa, 1995). Currency complementarity means that agents hold domestic and foreign money in fixed proportions, as no substitutability exists between them, and implies that the relative demands for domestic and foreign money do not depend on difference between the domestic and foreign interest rates.

[4] In CEE countries, the agents hold foreign money not only for foreign goods consumption, but also for domestic goods consumption. Thus, even if the preference for foreign money decreases continuously, a part of real estate or cars transactions are still performed using foreign currencies, especially the euro.

Dreger et al. (2007), Hsieh and Hsing (2009), and Fidrmuc (2009) investigate money demand in CEE countries. These empirical analyses, however, do not provide a theoretical framework. Therefore, we test the long-run relationship between money demand and its explanatory variables in a micro-founded model that generates two cointegrating equations. The first equation captures the sensitivity of real money demand for foreign currency to the opportunity cost spread of holding the money. The second equation captures the long-run relationship between real money demand for domestic currency and the opportunity cost of holding the domestic currency, the opportunity cost spread, and a scale variable.

We use Hansen's (1992) instability test to check for long-run relationships and the Dynamic OLS (DOLS) and the Fully-Modified OLS (FMOLS) methods to estimate the cointegrating relationships. We employ monthly data from 1999:M1 to 2015:M11 on four CEE countries -- Czech Republic, Hungary, Poland and Romania -- that use a floating exchange rate mechanism.[5] Our model proves, nevertheless, compatible with any exchange rate regime. Also, investigating the effect of currency substitution requires flexibility, but not necessarily a free-floating mechanism (e.g., Fidrmuc, 2009).

The rest of the paper is structured as follows. Section 2 briefly describes the models of money demand in an open economy. Section 3 presents our micro-founded money demand model. Section 4 parameterizes the model and identifies the cointegrating equations. Section 5 reports our empirical investigation for the CEE countries. Section 6 concludes.

---

[5] Croatia also has in place a managed floating exchange rate regime. However, until 2006, an exchange rate targeting regime was used. In addition, there are no data available for the monetary aggregate M2 for Croatia. Therefore, in order to have a consistent comparison of results for the CEE countries, we have decided to exclude Croatia from our sample.

## 2. The open-economy money demand models

Two different strands of literature characterize open-economy money demand models. The first strand considers micro-founded money demand models (e.g., Miles 1978, Bordo and Choudri 1982, and Hueng 1998, 2000). The second bulk of the literature empirically tests various money demand functions without considering the microeconomic foundations of their specifications (e.g., see early contributions of Cuddington 1983, and Leventakis 1993). Dreger et al. (2007), Hsieh and Hsing (2009), and Fidrmuc (2009) specifically examine money demand in CEE countries.

Miles (1978) uses Chetty's (1969) CES liquidity production function to derive the demand for domestic money relative to foreign money.[6] Bordo and Choudri (1982), however, argue that Miles (1978) misspecified his model, since he omitted income. In effect, Miles's portfolio choice model does not depend on the consumption-saving decision. Therefore, money demand does not depend on income. The money demand derives from the maximization of monetary service flows subject to an asset constraint. As a consequence, the ratio of domestic to foreign money demand depends only on the opportunity costs (i.e., domestic and foreign interest rates).

Bordo and Choudri (1982) derive money demand from a money-in-the-utility-function model. Their simplified model, however, is static, assuming that agents spend their entire income each period and that perfect interest rate arbitrage exists, thus, eliminating the effect of the exchange rate.

Hueng (1998, 2000) constructs cash-in-advance and shopping-time models to motivate money demand in an open economy. The cash-in-advance model in a two-country world, first studied by Stockman (1980), Lucas (1982) and Guidotti (1989), hinges on the assumption that agents purchase domestic and foreign goods with domestic and foreign currencies, respectively. The shopping-time model assumes that the time spent in purchasing domestic (foreign) consumption goods depends on the holdings of domestic (foreign) money. Thus, in the cash-in-advance and shopping-time models of Hueng (1998, 2000), foreign (domestic) money provides liquidity service only for foreign (domestic) good consumption. This critical assumption makes Hueng's (1998, 2000) models limited interest in economies where agents hold foreign money not only to purchase foreign goods, but also to purchase domestic goods. In addition, in some countries (i.e., the CEE countries), the agents can partially (or even totally) substitute an international money for their domestic money to purchase goods, regardless of the goods' origin.[7]

---

[6] Chen (1973) is a special case of Miles (1978), assuming a Cobb-Douglas demand function, which constrains the elasticity of substitution to equal one.

The second group of papers empirically examines whether currency substitution plays an important role in the demands for domestic and foreign money. Leventakis's (1993) two-country portfolio balance model shows that a change in the expected exchange rate affects the demand for domestic money by inducing its substitution with foreign money, which is the (direct) currency substitution effect, and with foreign bonds, which is the capital mobility effect.[8] If the exchange rate elasticity is high (i.e., if money demand is very sensitive to the exchange rate), this may indicate that currency substitution plays an important role in money demand.[9] If agents can switch between foreign and domestic money, then this may affect their money holdings.

Starting from this theoretical assumption, Dreger et al. (2007), Hsieh and Hsing (2009), and Fidrmuc (2009) examine the money demand in the CEE countries. Dreger et al. (2007)

---

[7] In our model, we make no distinction between foreign and domestic consumption goods. The representative agent's utility depends on the agent's entire consumption bundle, which mixes foreign and domestic goods, and on domestic and foreign money that produce liquidity services. The agent can invest in a portfolio composed of domestic and foreign money and bonds. By maximizing the (inter-temporal) utility function, we derive the money demand.

[8] The capital mobility effect is one of the two parts of the indirect currency substitution defined by McKinnon (1982). The second part is the substitution of domestic money with domestic bonds (under the assumption that uncovered interest parity holds, and a variation of the expected exchange rate induces a variation of the domestic interest rate).

[9] In Leventakis's (1993) general model, it is impossible to isolate the separate effects of currency substitution and capital mobility on the money demand. But if foreigners do not hold domestic currency assets, as for example in Cuddington (1983), it becomes possible to separate their effects.

study money demand in the new EU member states from 1995 to 2004. A well-behaved long-run money demand relationship exists only if the exchange rate appears as part of the opportunity cost. In the long-run cointegrating vector, the output elasticity exceeds unity. Over the entire sample, the exchange rate vis-à-vis the U.S. dollar proves significant and a more appropriate variable in money demand than the euro exchange rate.

Fidrmuc (2009) investigates the money demand with monthly data between 1994 and 2003 in six CEE countries (Czech Republic, Hungary, Poland, Romania, Slovakia, and Slovenia), and finds that the money demand depends significantly on the euro interest rate and on the exchange rate against the euro, which indicates possible instability of the money demand in these countries. The exchange rate elasticity, however, is low, which is, according to Fidrmuc, a good precondition for the eventual adoption of the euro by the CEE countries.[10] The euro area interest rates significantly shaped money demand in the CEE countries, indicating that capital mobility plays an important role in this region. The coefficient estimated for the euro area interest rate exceeds by a large amount the coefficients of domestic rates. Hsieh and Hsing (2009) find that the demand for M2 in Hungary positively associates with the nominal effective exchange rate and negatively associates with the deposit rate, the euro area interest rate, and the expected inflation rate from 1995-2005. They find an output elasticity near to unity, while Fidrmuc (2009) finds lower output elasticity, and a euro area interest rate coefficient higher than the domestic rate coefficient.

Elbourne and de Haan (2006) and Fidrmuc (2009) argue that a stable money demand and a transmission mechanism similar to that in the euro area will create good pre-conditions for the eventual introduction of the euro by new EU member states. Filosa (1995) and Dreger et al. (2007) also conclude that a stable money demand provides an important condition for using monetary aggregates in the conduct of monetary policy. Thus, these authors see that currency substitution indicates money demand instability. As such, a low exchange rate elasticity would, thus, provide a good pre-condition for the eventual adoption of the euro in the CEE countries.

---

[10] He also finds that the parameters of money demand in CEE countries closely approximates those in developed countries, which gives a good pre-condition for euro adoption.

On the contrary, this point of view is problematic. In fact, currency substitution results from monetary integration. Miles (1978), McKinnon (1982), Bordo and Choudri (1982), Leventakis (1993) and Hueng (1998) remind us that if people's money holdings change with foreign monetary developments, such as the foreign interest rate and the exchange rate, then the isolation mechanism of the floating exchange rate system will not work, thus providing the policymaker lower control from stabilization policies. Currency substitution does reduce the stability of the money demand in each country, but this does not mean that the global money demand is less stable. In fact, while defining meaningful monetary aggregates, McKinnon (1982) suggests that currency substitution makes an appropriately defined global (monetary union) money supply rather than national money supplies more relevant for studying global (union) inflation. Thus, when currency substitution occurs, it then becomes more appropriate to conduct a global (union) monetary policy rather than a national monetary policy. From this point of view, currency substitution between CEE currencies and the euro gives a signal of monetary integration between the two areas and a good pre-condition for the eventual adoption of the euro by the CEE countries.

Our research relies on both strands of literature described above. First, we propose a micro-founded money demand model, which separates the currency substitution effect from the money demand sensitivity to exchange rates. Second, we parameterize the model and we empirically investigate the long-run money demand with an application to four CEE countries.

## 3. The micro-founded open economy money demand model

The domestic agent living in an outlying country (i.e., CEE country) orders his preferences according to the lifetime utility function:

$$V_t = E_t\left[\sum_{i=0}^{\infty}\beta^i U\left(\frac{X_{t+i}}{P_{t+i}}, \frac{M_{t+i}}{P_{t+i}}, \frac{S_{t+i}M^*_{t+i}}{P_{t+i}}\right)\right], \quad (1)$$

where $X_t$ is monetary consumption spending measured in terms of domestic money, $P_t$ is the price index, $M_t$ is domestic money holding, and $M^*_t$ is foreign money holding. If one unit of foreign money equals $S_t$ units of domestic money, then $S_t M^*_t$ equals the domestic money value of the domestic agent's foreign money holding. The expectations operator $E_t[.]$ is conditional on the information available at time t.

The agent faces the following budget constraint:

$$M_{t-1}(1-\phi) + S_t M^*_{t-1}(1-\phi) + B_{t-1}(1+i_t) + S_t B^*_{t-1}(1+i^*_t) + Z_t = X_t + M_t + S_t M^*_t + B_t + S_t B^*_t,$$

where $B_t$ is the monetary value (in terms of domestic money) of domestic bond holding, and $B^*_t$ is the monetary value (in terms of foreign money) of foreign bond holding, which is an imperfect substitute for the domestic bonds because of exchange rate risk; $Z_t$ is the lump-sum monetary transfer to the agent from the government; and $i_{t+1}$ and $i^*_{t+1}$ are the nominal domestic and foreign interest rates. As bonds are nominally risk-free, $i_{t+1}$ and $i^*_{t+1}$ are known at time t.

The parameter $\phi$ represents the cost the agent faces for holding money. We model this cost as a proportional cost to simplify the analysis. It stands for the charges related to the use of a bank account, the cost of a bank card, the renting of a bank safe deposit box, and the cost of cash theft or loss. In standard money demand models, the proportional cost is neglected (i.e., $\phi = 0$); hence, the interest rate cannot be negative. Assuming a non-zero $\phi$ addresses in a simple way negative interest rates (a similar approach is adopted by Benati et al., 2016).

We calculate real consumption spending from the budget constraint as follows:

$$\frac{X_t}{P_t} = \frac{M_{t-1}}{P_{t-1}}\frac{P_{t-1}}{P_t}(1-\phi) + \frac{S_t M^*_{t-1}}{P_{t-1}}\frac{P_{t-1}}{P_t}(1-\phi) + \frac{B_{t-1}}{P_{t-1}}\frac{P_{t-1}}{P_t}(1+i_t) + \frac{S_t B^*_{t-1}}{P_{t-1}}\frac{P_{t-1}}{P_t}(1+i^*_t) + \frac{Z_t}{P_t} \\ - \frac{M_t}{P_t} - \frac{S_t M^*_t}{P_t} - \frac{B_t}{P_t} - \frac{S_t B^*_t}{P_t}. \qquad (2)$$

The agent maximizes equation (1) with respect to $\frac{M_t}{P_t}, \frac{S_t M^*_t}{P_t}, \frac{B_t}{P_t}$ and $\frac{S_t B^*_t}{P_t}$ subject to equation (2). Let $U_H$ denote the partial derivative of U with respect to H. The first-order conditions are as follows:

$$E_t\left[\frac{\partial V_t}{\partial \frac{M_t}{P_t}}\right] = 0 \Leftrightarrow E_t\left[-U_{\frac{X_t}{P_t}} + \beta U_{\frac{X_{t+1}}{P_{t+1}}}\frac{P_t}{P_{t+1}}(1-\phi) + U_{\frac{M_t}{P_t}}\right] = 0. \qquad (3)$$

$$E_t\left[\frac{\partial V_t}{\partial \frac{S_t M^*_t}{P_t}}\right] = 0 \Leftrightarrow E_t\left[-U_{\frac{X_t}{P_t}} + \beta U_{\frac{X_{t+1}}{P_{t+1}}}\frac{S_{t+1}}{S_t}\frac{P_t}{P_{t+1}}(1-\phi) + U_{\frac{S_t M^*_t}{P_t}}\right] = 0. \qquad (4)$$

$$E_t\left[\frac{\partial V_t}{\partial \frac{B_t}{P_t}}\right] = 0 \Leftrightarrow E_t\left[-U_{\frac{X_t}{P_t}} + \beta U_{\frac{X_{t+1}}{P_{t+1}}}\frac{P_t}{P_{t+1}}(1+i_{t+1})\right] = 0. \qquad (5)$$

$$E_t\left[\frac{\partial V_t}{\partial \frac{S_t B^*_t}{P_t}}\right] = 0 \Leftrightarrow E_t\left[-U_{\frac{X_t}{P_t}} + \beta U_{\frac{X_{t+1}}{P_{t+1}}}\frac{S_{t+1}}{S_t}\frac{P_t}{P_{t+1}}(1+i^*_{t+1})\right] = 0. \qquad (6)$$

Equations (4), (5), and (6) describe the direct and indirect currency substitution. First, consider equation (4). This equation generates the optimal foreign money holding. It assumes that direct currency substitution exists. On the contrary, imagine that the agent cannot substitute foreign and domestic currencies. Then the agent cannot choose his level of foreign money holding, which is fixed: $\frac{S_t M_t^*}{P_t} = \frac{S_t M_{t-1}^*}{P_t}$, or in a more general way, $\frac{S_t M_t^*}{P_t} = \frac{S_t (M_{t-1}^* + em_t^*)}{P_t}$, where $em_t^*$ is exogenous. In any case, if no currency substitution exists, the agent cannot decide the level of $\frac{S_t M_t^*}{P_t}$ and he cannot optimize his utility function with respect to it. Therefore, equation (4) does not hold if currency substitution does not exist. Equation (4) then is a consequence of currency substitution.[11]

Next, consider equation (6). This equation produces the optimal foreign bond holding. It assumes capital mobility. If international capital flows are restricted, then the agent cannot choose his foreign bond holding, which is fixed in the extreme case: $\frac{S_t B_t^*}{P_t} = \frac{S_t B_{t-1}^*}{P_t}$, or in a more general way, $\frac{S_t B_t^*}{P_t} = \frac{S_t (M_{t-1}^* + eb_t^*)}{P_t}$, where $eb_t^*$ is exogenous. If capital mobility does not exist, then the agent cannot determine $\frac{S_t B_t^*}{P_t}$ and equation (6) does not hold. Equation (6) then is a consequence of capital mobility.

Finally, consider equation (5). This equation makes the optimal domestic bond holding. Indirect currency substitution assumes that the agent can freely choose domestic bond holding. If the agent cannot determine $\frac{B_t}{P_t}$, then equation (5) does not hold.

---

[11] The ownership of foreign money by residents is not proof of (direct) currency substitution. Rather, the responsiveness of foreign money demand to the exchange rate or to the foreign interest rate provides clear evidence of currency substitution.

Now, suppose the absence of currency substitution, direct or indirect, which corresponds to the hypothesis of exogeneity of $\frac{S_t M_t^*}{P_t}$, $\frac{S_t B_t^*}{P_t}$, and $\frac{B_t}{P_t}$, and equations (4), (5), and (6) do not hold. The agent only determines the optimal domestic money holding, hinging on equation (3), the only equation that holds. Equation (3) shows a relationship between the current and one-period ahead marginal utility of consumption, foreign and domestic real cash balances, and the inflation rate. As the various marginal utilities depend on $\frac{X_t}{P_t}$, $\frac{M_t}{P_t}$ and $\frac{S_t M_t^*}{P_t}$, equation (3) indicates that the domestic money demand $\frac{M_t}{P_t}$ depends on $\frac{X_t}{P_t}$, $\frac{S_t M_t^*}{P_t}$, $\frac{X_{t+1}}{P_{t+1}}$, $\frac{M_{t+1}}{P_{t+1}}$, $\frac{S_{t+1} M_{t+1}^*}{P_{t+1}}$ and $\frac{P_t}{P_{t+1}}$. For this somewhat complex formulation of the open-economy money demand, we observe that the one-period ahead exchange rate $S_{t+1}$ enters the money demand function. So, even after removing any possibility of currency substitution, the money demand depends on the exchange rate, a result that contrasts with the whole literature devoted to currency substitution.

The intuition behind this result is simple. In effect, the variation of the exchange rate influences the liquidity services provided by foreign money holding. Even if the agent will not replace domestic money with foreign money (or with domestic or foreign bonds) in response to changes in their relative rate of return, the agent can switch between consumption and domestic money holding to respond to liquidity shocks caused by the exchange rate change.

Therefore, the agent responds to the exchange rate change by modifying domestic money holding. Consequently, in any case, we conclude that a non-zero exchange rate elasticity results from direct or indirect currency substitution.[12]

Thus, if we remove any possibility of currency substitution, equation (3) shows that the domestic money demand depends on the exchange rate and the inflation rate. The money demand, however, does not depend on the domestic and foreign interest rates. If we make the assumption, which seems realistic, that the agent controls domestic bond holdings, then equation (5) holds too. Multiplying equation (3) by $(1+i_{t+1})/(1-\phi)$ and subtracting equation (5) gives:

$$-U_{\frac{X_t}{P_t}}(i_{t+1}+\phi)+U_{\frac{M_t}{P_t}}(1+i_{t+1})=0. \tag{7}$$

No more role exists for risky variables in equation (7), in particular reference to the inflation rate or the exchange rate. Equation (7) shows that domestic money demand depends on $\frac{X_t}{P_t}, \frac{S_t M_t^*}{P_t}$, and the domestic interest rate $i_{t+1}$.

We can write the money demand function in many ways. We can use equation (3) to express money demand as a complex function involving the exchange rate and the inflation rate, or we can use a mix of equations (3) and (5) express money demand in a way that excludes these two variables. If we add the hypothesis that the agent controls foreign bond holdings, which assumes capital mobility, equation (6) also holds. Multiply equation (6) by $i_{t+1}+\phi$ and then subtracting equation (7) gives:

$$U_{\frac{M_t}{P_t}}(1+i_{t+1})-E_t\left[\beta U_{\frac{X_t}{P_{t+1}}}\frac{S_{t+1}}{S_t}\frac{P_t}{P_{t+1}}(1+i_{t+1}^*)(i_{t+1}+\phi)\right]=0$$

---

[12] In fact, the consumption – money substitution effect that we describe is possibly more important than the currency substitution effect, depending on the value of the liquidity elasticity defined hereafter.

This equation leads to a complex, intractable formulation of the domestic money demand depending on all variables considered in the model, the exchange rate, the inflation rate, and the foreign and domestic interest rates.

Finally, we add the assumption of direct currency substitution, which means that equation (4) holds. Then multiplying equation (4) by $(1+i^*_{t+1})/(1-\phi)$ and subtracting equation (6) gives:

$$-U_{\frac{X_t}{P_t}}(i^*_{t+1}+\phi)+U_{\frac{S_t M^*_t}{P_t}}(1+i^*_{t+1})=0 \qquad . \tag{8}$$

Both equations (7) and (8) depend on known (non-random) terms that express the domestic money demand as a function of domestic and foreign interest rates (but independent of the exchange rate and the inflation rate).

To conclude, a non-zero exchange rate elasticity does not prove that currency substitution exists. Indeed, a consumption–money substitution effect also influences the money demand and, thus, the sensitivity of money demand to international variables should not depend solely on currency substitution. And even if currency substitution exists, we can still express the money demand as a function independent of the exchange rate.

Finally, we cannot test the assumption of currency substitution in a model that depends crucially on this hypothesis. If equations (3) to (6) hold, then we assume indirect and direct currency substitution, which is a core hypothesis of the model and which we cannot test. Fortunately, this assumption is not as strict as it seems, as the micro-founded model permits a flexible degree of substitution, which could be higher or lower, consistent with a high degree of currency substitution or with currency complementarity. We investigate the degree of currency substitution between the euro and the currencies of the CEE countries and estimate the intensity of the parameters that explain the sensitivity of money demand to international economic variables.

## 4. A parameterization of the utility function

Following Miles (1978), we parameterize our model by specifying that the domestic and foreign currency enter a CES liquidity production function $L_t/P_t$, and that the produced liquidity and real consumption also enter a CES function:

$$U = \left\{ \theta \left( \frac{X_t}{P_t} \right)^\eta + (1-\theta) \left( \frac{L_t}{P_t} \right)^\eta \right\}^{\frac{1}{\eta}}, \text{ with } \eta = \frac{\zeta-1}{\zeta}, \tag{9}$$

where $\zeta = 1/(1-\eta)$ is the elasticity of substitution between consumption and liquidity, and

$$\frac{L_t}{P_t} = \left\{ \delta \left( \frac{M_t}{P_t} \right)^\gamma + (1-\delta) \left( \frac{S_t M_t^*}{P_t} \right)^\gamma \right\}^{\frac{1}{\gamma}} \text{ with } \gamma = \frac{\sigma-1}{\sigma}, \tag{10}$$

where $\sigma = 1/(1-\gamma)$ is the elasticity of substitution between domestic and foreign money in the liquidity production function.

In the case of a zero elasticity of substitution $\sigma$ $(\zeta)$, the CES function becomes the Leontief function, which indicates that domestic and foreign money (or consumption and liquidity) are perfect complements. In the particular case of a unitary elasticity of substitution, the CES function becomes a Cobb-Douglas function.

When the elasticity of substitution $\sigma$ $(\zeta)$ increases, it is easier to replace one currency with another (or to replace consumption with liquidity). In the extreme perfect substitution case, the elasticity of substitution goes to infinity. A value of $\sigma > 1$ $(\zeta > 1)$ indicates substitutability between domestic and foreign moneys (between consumption and liquidity), while a value $\sigma < 1$ ($\zeta < 1$) indicates complementarity between them. If we confirm the assumption of CEE countries' monetary integration with the euro area, then these currencies must be highly substitutable with the euro. Therefore, we must pay particular attention to the $\sigma$ ($\zeta$) estimation.

The term $\left\{\delta\left(\frac{M_t}{P_t}\right)^\gamma + (1-\delta)\left(\frac{S_t M_t^*}{P_t}\right)^\gamma\right\}^{\frac{1}{\gamma}}$ represents the liquidity production function whose inputs are domestic and foreign money holdings, and where $\delta$ is the share parameter. The condition $\delta > 0.5$ ($\delta < 0.5$) means that the domestic money is more (less) liquid than the euro in the eyes of the CEE countries' representative agent. The CES liquidity production function and the real consumption are next combined according to a CES utility function[13] where $\theta$ is the share parameter. We restrict the parameters of the utility and liquidity production functions so that $0 < \theta < 1$, $0 < \delta < 1$, $0 \leq \sigma < +\infty$ and $0 \leq \zeta < +\infty$.

Calculating the partial derivatives $U_{\frac{X_t}{P_t}}, U_{\frac{M_t}{P_t}}$, and $U_{\frac{S_t M_t^*}{P_t}}$, and inserting them successively into equations (7) and (8) gives:

$$\frac{1-\theta}{\theta}\left(\frac{L_t}{P_t}\right)^{\eta-\gamma}\left(\frac{X_t}{P_t}\right)^{1-\eta} = \frac{1}{\delta}\frac{i_{t+1}+\phi}{1+i_{t+1}}\left(\frac{M_t}{P_t}\right)^{1-\gamma}, \tag{11}$$

and

$$\frac{1-\theta}{\theta}\left(\frac{L_t}{P_t}\right)^{\eta-\gamma}\left(\frac{X_t}{P_t}\right)^{1-\eta} = \frac{1}{1-\delta}\frac{i_{t+1}^*+\phi}{1+i_{t+1}^*}\left(\frac{S_t M_t^*}{P_t}\right)^{1-\gamma}. \tag{12}$$

As the left-hand terms of equations (11) and (12) are the same, equations (11) and 12 leads to:

$$\frac{S_t M_t^*}{P_t} = \left[\frac{1-\delta}{\delta}\left(\frac{i_{t+1}+\phi}{1+i_{t+1}} \Big/ \frac{i_{t+1}^*+\phi}{1+i_{t+1}^*}\right)\right]^\sigma \frac{M_t}{P_t}. \tag{13}$$

---

[13] The generalized utility function $\frac{1}{1-\alpha}U^{1-\alpha}$ leads to the same money demand function as the CES U function, which does not depend on the risk aversion parameter $\alpha$. Therefore, we ignore risk aversion, as it does not affect the money demand equation.

We denote $oc_{t+1} = \frac{i_{t+1} + \phi}{1 + i_{t+1}}, oc_{t+1}^* = \frac{i_{t+1}^* + \phi}{1 + i_{t+1}^*}$ as the opportunity costs of holding domestic and foreign moneys. If $\phi = 0$, then these opportunity costs equal the discounted interest rates.[14]

We rewrite equation (13) in natural logarithms as follows:

$$\ln\left(\frac{M_t}{S_t M_t^*}\right) = \sigma \ln\left(\frac{\delta}{1-\delta}\right) - \sigma\left(\ln oc_{t+1} - \ln oc_{t+1}^*\right). \tag{14}$$

Equation (14) is similar to the relationship between domestic and foreign money demand derived by Miles (1978), except that we replace the terms $1 + i_{t+1}$ and $1 + i_{t+1}^*$ in Miles (1978) with $oc_{t+1}$ and $oc_{t+1}^*$. By integrating the liquidity production function and the consumption in a CES function, we demonstrate that the Bordo-Choudri (1982) criticism of Miles's (1978) money demand equation is not relevant, as the equation does not reflect the omission of consumption or income. When we analyze domestic money demand compared to foreign money demand, we do not need to add a scale variable such as consumption or income. We only need the share parameter and the elasticity of substitution to explain the shift in domestic money demand relative to foreign money demand.

If $\ln(M_t / S_t M_t^*)$ and $(\ln oc_{t+1} - \ln oc_{t+1}^*)$ are I(1), then equation (14) describes a cointegrating relationship.[15] In the long run, equation (14) holds exactly, and so it appears as a long-run money demand equation. In the short run, however, because adjustment takes time, a temporary disequilibrium $\varepsilon_t$ exists such that the relationship is:

$$\ln\left(\frac{M_t}{S_t M_t^*}\right) = \kappa_0 - \kappa_1\left(\ln oc_{t+1} - \ln oc_{t+1}^*\right) + \varepsilon_t, \tag{15}$$

---

[14] The empirical money demand literature depicts the interest rate as the opportunity cost of money holding, which assumes that $\phi = 0$, and uses it as a regressor in the money demand equation. But, as the interest rate is perceived one period later, we must discount it to the present, and the discounted interest rate enters the money demand regression.

[15] Standard unit-root tests and panel unit-root tests confirm that the variables involved in equation (17) are I(1) processes (see Dreger et al. 2007; Fidrmuc, 2009; Hsieh and Hsing, 2009).

where the elements $\kappa_0$ and $\kappa_1$ of the cointegrating vector related to the structural parameters by $\kappa_0 = \sigma \ln[\delta/(1-\delta)]$ and $\kappa_1 = \sigma$.

The relative money demand function in equation (15) is not the only result of the model. Our model also delivers a money demand equation dependent on a scale variable (consumption or income), more consistent with the standard empirical money demand equations estimated in the literature.

Returning to equation (11) expressed as:

$$\frac{M_t}{P_t} = \left\{ \delta \left( \frac{1-\theta}{\theta} \right) \frac{1+i_{t+1}}{i_{t+1}+\phi} \right\}^{\frac{1}{1-\eta}} \left\{ \delta + (1-\delta) \left( \frac{S_t M_t^*}{P_t} \bigg/ \frac{M_t}{P_t} \right)^{\gamma} \right\}^{\frac{\eta-\gamma}{(1-\eta)\gamma}} \left( \frac{X_t}{P_t} \right), \qquad (16)$$

and insert equation (13) into equation (16) to get:

$$\frac{M_t}{P_t} = \left( \frac{1-\theta}{\theta} \right)^{\zeta} oc_{t+1}^{-\zeta} \delta^{\frac{\sigma(\zeta-1)}{\sigma-1}} \left\{ 1 + \psi \left( \frac{oc_{t+1}}{oc_{t+1}^*} \right)^{\sigma-1} \right\}^{\frac{\zeta-\sigma}{\sigma-1}} \left( \frac{X_t}{P_t} \right), \text{ with } \psi = \left( \frac{1-\delta}{\delta} \right)^{\sigma}. \qquad (17)$$

Consider that the country in question is an outlying country, and its currency offers no liquidity services to foreign agents. In this case, we assume that the money of the foreign country (i.e., the euro) is an international money that offers liquidity services to the agent of the outlying country, but not the reverse. The foreign agent does not demand money from the outlying country and the total demand for money of this country simply equals $M_t/P_t$.

Taking the logarithm of equation (17) gives:

$$\ln\left(\frac{M_t}{P_t}\right) = \zeta \ln\left(\frac{1-\theta}{\theta}\right) + \frac{\sigma(\zeta-1)}{\sigma-1}\ln\delta - \zeta \ln oc_{t+1} + \left(\frac{\zeta-\sigma}{\sigma-1}\right)\ln\left\{1+\psi\left(\frac{oc_{t+1}}{oc_{t+1}^*}\right)^{\sigma-1}\right\} + \ln\left(\frac{X_t}{P_t}\right) \quad (18)$$

We observe that the money demand equation (18) conforms to the standard result of a unitary output elasticity. The model is a somewhat complex nonlinear equation and we decide to restrict our analysis to a simplified linearized version similar to those in empirical money studies.[16]

If we assume that $\ln oc_{t+1} - \ln oc^*_{t+1}$ is approximately a constant $s$, which we consider as a long-run spread, then the Taylor expansion yields:

$$\ln\left(\frac{M_t}{P_t}\right) = \zeta \ln\left(\frac{1-\theta}{\theta}\right) + \left(\frac{\zeta-\sigma}{\sigma-1}\right)\left[\ln\{1+\psi\exp[(\sigma-1)s]\} - \frac{s\psi(\sigma-1)\exp[(\sigma-1)s]}{1+\psi\exp[(\sigma-1)s]}\right]$$
$$+ \frac{\sigma(\zeta-1)}{\sigma-1}\ln\delta - \zeta\ln oc_{t+1} + \frac{\psi(\zeta-\sigma)\exp[(\sigma-1)s]}{1+\psi\exp[(\sigma-1)s]}(\ln oc_{t+1} - \ln oc^*_{t+1}) + \ln\left(\frac{X_t}{P_t}\right) \quad (19)$$

If $\ln(M_t/P_t), \ln oc_{t+1}, \ln oc_{t+1} - \ln oc^*_{t+1}$, and $\ln(X_t/P_t)$ are I(1), then equation (19) describes a second cointegrating relationship. In the long run, this money demand equation holds exactly, but in the short run, the money depends also on a stationary disequilibrium $\eta_t$:

$$\ln\left(\frac{M_t}{P_t}\right) = \omega_0 - \omega_1 \ln oc_{t+1} + \omega_2(\ln oc_{t+1} - \ln oc^*_{t+1}) + \omega_3 \ln\left(\frac{X_t}{P_t}\right) + \eta_t, \quad (20)$$

where the elements of the cointegrating vector relate to the structural parameters by

$$\omega_0 = \zeta \ln\left(\frac{1-\theta}{\theta}\right) + \frac{\sigma(\zeta-1)}{\sigma-1}\ln\delta + \left(\frac{\zeta-\sigma}{\sigma-1}\right)\left[\ln\{1+\psi\exp[(\sigma-1)s]\} - \frac{s\psi(\sigma-1)\exp[(\sigma-1)s]}{1+\psi\exp[(\sigma-1)s]}\right],$$
$$\omega_1 = \zeta, \omega_2 = \frac{\psi(\zeta-\sigma)\exp[(\sigma-1)s]}{1+\psi\exp[(\sigma-1)s]}, \text{ and } \omega_3 = 1.$$

The parameter $\omega_1$ measures the interest rate elasticity and the sign of the parameter $\omega_2$ depends on the difference between the two elasticities of substitution $\zeta$ and $\sigma$.

The money demand equations (15) and (20) follow in the line of Meltzer (1963), or fit the category of Baumol-Tobin models (i.e., inventory-theoretic models, Baumol 1952 and Tobin 1956), which consider, as explanatory variables, the log of opportunity cost, and not the opportunity cost itself, as in Cagan's (1956) approach.

---

[16] Hueng (2000) made the same choice when confronted with a nonlinear money demand function.

Cagan's (1956) semi-log form is the most commonly used specification for empirical analysis of money demand, where the opportunity cost generally becomes the interest rate. The existence of low, or even negative, interest rates makes it difficult or impossible to use a log-log money demand specification when we $\phi = 0$. But, for a high enough value of $\phi$, or for much higher interest rates, the log-log money demand specification offers an interesting alternative.[17]

**5. An application to the CEE countries**

W can apply the model described in the previous sections to the CEE countries, where an international currency (the euro) offers liquidity services to domestic agents. In the CEE countries, the ratio of foreign currency deposits to total deposits is high, and most foreign currency deposits are euros.

We use monthly statistics for the Czech Republic, Hungary, Poland and Romania, to test the two long-run money demand equations -- equations (15) and (20) -- from 1999M1 to 2015M11. The data come from the International Financial Statistics (IFS) database, the Eurostat database, the OECD database, and statistics provided by the national central banks.

In equation (15), the ratio of domestic to foreign currency deposits proxies for the money demand in our model, as the structure of money in circulation (cash) is unknown. For equation (20), we use M2 to compute real money demand, as researchers commonly use broad money to estimate money demand in the CEE countries. For robustness purpose, in equation (20) we also employ M1 in equation (20) to measure real money demand. We use the money market rate, the consumer price index, and household consumption expenditure.[18] A complete data description appears in the Appendix.

---

[17] Lucas (2000) compares the two types of money demand and expresses a preference for the log-log form (for additional arguments for the log-log specification see, also, Benati et al, 2016 and Miller, Martins, and Gupta forthcoming). In contrast, Ireland (2009) argues for the semi-log form. The debate on the best choice of the form for the money demand equation still continues.

[18] Because our model relies on monetary consumption spending, we have retained household consumption expenditure for the scale variable in equation (20). This variable is available on a quarterly basis only, and we

For the opportunity cost ϕ, we do not distinguish between money in circulation and deposits. Thus, the proportional cost is identical for all money substitutes. In addition, we consider that the value of ϕ is sufficiently small (see previous studies). In this context, we set for monthly data a value of 0.00082953 (equivalent to 1% on an annual basis), which corresponds to a loss rate or negative return of 1% for narrow money[19] advanced by Lucas and Nicolini (2015) and Benati et al. (2016). As explained by Benati et al. (2016), a non-zero (strictly positive) value for ϕ is a necessary assumption when considering a log-log money demand, especially if the model is applied to data containing periods of nearly zero interest rates.

We estimate the cointegration equations (15) and (20) by the DOLS method of Saikkonen (1991) and Stock and Watson (1993) and by the FMOLS method of Phillips and Hansen (1990). Both procedures produce asymptotically unbiased estimators even in the absence of strong exogeneity of the regressors.

Both methods are also consistent with the triangular representation of Phillips (1988, 1991) of cointegrated I(1) processes. This representation is valid for any cointegrating rank, but we assume that a cointegrating rank of 2. Consider a n-vector $Y_t = (y_{1t} \ y_{2t} \ Y_{3t}')'$, where $y_{1t}$ and $y_{2t}$ are one-dimensional I(1) processes and $Y_{3t}$ is a (n-2)-dimensional I(1) process. Assume that $Y_t$ is cointegrated with rank 2. The triangular representation is an n-equations system consisting of two cointegrating regressions:

$$\begin{cases} y_{1t} = \mu_1 + a_1' Y_{3t} + z_{1t} \\ y_{2t} = \mu_2 + a_2' Y_{3t} + z_{2t} \end{cases}, \quad (21)$$

and a (n-2)-dimensional I(1) process:

---

use a cubic spline function to generate the monthly frequency. In line with previous empirical estimations, however, we also use the industrial production index on a monthly basis for the scale variable. Using the Census X13, we seasonally adjusted both variables.

[19] For cash, it is likely that the value of ϕ increases to 2%. This is the estimate by Alvarez and Lippi (2009) of the probability of cash theft in Italy, for example. A value of 1%, however, seems *a priori* more appropriate to describe the loss of cash and costs associated with owning a bank account.

$$\Delta Y_{3t} = \mu_3 + Z_{3t}, \tag{22}$$

where $Y_{3t}$ is not cointegrated and $(z_{1t} \ z_{2t} \ Z_{3t}^{'})^{'}$ is a zero-mean stationary process.

This triangular representation presents $y_{1t}$ and $y_{2t}$ as the "dependent" variables, but in fact, it does not require that they are the only endogenous variables, as the hypothesis of strong exogeneity of the (n-2)-dimensional regressor $Y_{3t}$ is not required. Note that this representation assumes that only one "dependent" variable exists in each of the cointegrating regressions and that an equation for each "dependent" variable also exists. Equations (15) and (20) prove consistent with this triangular representation with $y_{1t} = \ln(M_t/S_t M_t^*)$, $y_{2t} = \ln(M_t/P_t)$ and $Y_{3t} = (\ln oc_{t+1} \ \ln oc_{t+1} - \ln oc_{t+1}^* \ \ln(X_t/P_t))^{'}$. The DOLS and FMOLS estimators of system (21) are asymptotically equivalent to the Johansen's maximum likelihood estimation method (Johansen, 1988), based on the vector error-correction model. They deliver standard statistics (e.g., t- and Wald-statistics) that are asymptotically normally distributed.

Before estimating the long-run relationship, however, we want to ensure that our series are I(1). Therefore, in the first step, we apply ADF and PP unit root tests, including a constant term. Table 1 presents the results and show that our variables are I(1).[20] Therefore, we proceed with the cointegration analysis for both equations (15) and (20).

For each equation, we test the existence of a long-run relationship based on Hansen's (1992) instability test, relying on the $L_c$ statistic. The null hypothesis of this cointegration test is the presence of cointegration. For the DOLS-type estimations, we choose the number of leads and lags using the Akaike information criteria, while for the FMOLS (with Bartlett kernel), we use a Newey-West automatic bandwidth rule. We test different hypotheses on the parameters based on the Wald test t-statistic.

---

[20] A small exception occurs for the log of the real industrial production in Romania.

The results for each equation (15) and (20) appear in Tables 2, 3, and 4. First, looking at equation (15), we notice that in almost all the cases the results of the cointegration test differ between the DOLS and FMOLS estimations. The cointegration exists for the DOLS estimation only. Hungary represents an exception, as the $L_c$ statistic shows the existence of a long-run relationship for both the DOLS and FMOLS estimations. Now, if we accept the hypothesis of cointegration, both methods show that the elasticity of substitution is low, less than 1, and positive (except for Poland). In addition, the Wald t-statistic shows that $\kappa_1 \neq 1$. We see that, according to this criterion, the monetary integration of CEE countries with the euro area is reduced, as the value of elasticity between the currencies estimated by $\kappa_1$ indicates currency complementary rather than currency substitution. This affirmation, however, is true to a smaller extent for Romania, where a many current transactions use euros.[21]

The results for our second cointegrating relationship in equation (20) appear in Table 3. A first set considers real household consumption as a scale variable, while a second set considers real industrial production as a scale variable.

Several conclusions emerge from these findings. First, the coefficients' sign and significance level reports, in general, a strong correspondence between the DOLS and FMOLS estimations. We can validate the cointegration relationship, however, in all cases only for the DOLS approach. The FMOLS estimation exhibit more mitigated findings, as we reject the null hypothesis of cointegration 3 of the 8 cases.

Second, the interest elasticity $\omega_1$ shows the expected sign and it is significant, except for Poland when consumption is the scale variable. Its low value indicates that consumption and liquidity are complements, not substitutes, except for Romania with industrial production as a scale variable. The positive sign of the spread's coefficient $\omega_2$ implies that $\sigma < \zeta$, which reflects the low value of $\sigma$, in agreement with the estimates of equation (20). Third, the coefficient $\omega_3$ is positive, as expected, in all the cases for household consumption and with one exception (Romania) for industrial production.[22] The consumption and output elasticities, however, exceed 1. We, thus, reject the hypothesis of a unitary elasticity.

---

[21] According to the National Bank of Romania statistics (monthly bulletins), the ratio of foreign currency to total deposits over 1999-2015 exceeds 50%, decreasing from 70% at the beginning of the 2000s, to 34% in the present.

If we compare the CEE countries, the Hungarian and Romanian money demand responds more to the opportunity cost based on the internal discounted interest rate, while the Czech money demand responds more to the opportunity cost spread. In addition, real output exhibits greater importance for Czech and Polish money demand compared with Hungarian and Romanian money demand. All in all, the small elasticity of substitution may imply less monetary integration for CEE countries.

We check the robustness of these findings, using M1 instead of M2 for the money demand in equation (20). Table 4 presents the results. As in the previous case, a string correspondences exist between the DOLS and FMOLS estimates. Only for the DOLS approach does the Hansen's (1992) instability test shows the existence of a cointegrating relationship for all countries. The FMOLS estimator rejects the null of cointegration in 5 of the 8 cases. At the same time, $\omega_1$ exhibits the expected sign and is significant in all cases, except Romania, when consumption is the scale variable. The spread of opportunity cost ($\omega_2$) positively affects money demand, indicating that $\sigma < \zeta$, with two exceptions, Hungary and Poland for the DOLS estimation with household consumption as the scale variable (this result does not stand when we use industrial production). A slight difference in results occurs for M2 with industrial production. The FMOLS estimate for Poland finds that $\omega_3$ is negative.

All in all, our empirical estimates fit the theoretical assumptions synthetized in equations (15) and (20). The employed tests confirm, in general, the long-run relationship, explaining the money demand in CEE countries. Consequently, money demand in CEE countries depends on the opportunity cost of holding the money (i.e., the discounted money market rate), the spread of the opportunity cost, and the scale variable (i.e., consumption or output).

---
[22] This result may reflect the fact that real industrial production for Romania is not an I(1) process.

## 6. Conclusions

We investigated the money demand in CEE countries starting from a theoretical model with micro-foundations, which incorporates both the currency substitution and money demand sensitivity to exchange rate effects. This model established a channel for an exchange rate effect on money demand, even absent currency substitution. We apply this model to CEE countries, where the euro offers liquidity services to domestic agents, while money of CEE countries does not offer liquidity service to residents of the euro area. .

The model parameterization shows that CEE money demand includes two complementary cointegrating relationships, which represent an original result of our model. The empirical findings revealed by Hansen's (1992) instability test, on the one hand, and the DOLS and FMOLS estimators, on the other hand, document the two cointegrating relationships, where real CEE money demand depends on the opportunity cost of holding the money as well as real consumption or real output. A consensus exists between the DOLS and FMOLS results, and the findings are robust to the use of M2 or M1 for assessing the money demand in equation (20). In general, the CEE countries' agents perceive that domestic currency is more liquid than the euro and a low level of substitution exists between domestic currencies and the euro. Previous empirical studies on CEE countries' money demand test the money demand sensitivity to international factors and report, in general, a high substitution level. Our micro-founded model shows a lower level of substitution and complementarity, not substitutability, between CEE currencies and the euro. Therefore, we should view the high degree of substitution between CEE currencies and the euro reported in prior studies with caution, because these studies do not consider exchange rate effects on money demand in the absence of currency substitution. Also, they do not consider, as possible, complementary between CEE currencies and the euro.

Our empirical results, however, do not represent incontestable proof of two long-run money demand relationships, as the results prove less robust for the FMOLS estimator. Nevertheless, the empirical findings clearly show a reduced degree of substitution. Thus, the monetary integration of CEE countries with the euro area seems lower for the moment. This result, however, depends on stronger confidence in the CEE domestic currencies, and by the increased liquidity service they provide. Moreover, other criteria such as the adoption of EU regulations and the level of financial integration, show that the CEE countries are more and more prepared for euro adoption.

The policy implications of our study are twofold. We show that the monetary authorities in the CEE countries should consider, in the money demand estimation, not only the opportunity cost of holding the money and consumption or output, but also the effect of the exchange rate, which occurs even absent strong currency substitution. At the same time, we posit that the degree of substitution between CEE currencies and the euro should not imply *per se* reduced monetary integration. Actual macroeconomic policies increased the confidence of CEE agents in domestic currencies. These countries need the political will to join the euro zone, even though efforts must continue toward higher monetary integration.

# References

Alvarez, F., and F., Lippi, (2009). Financial innovation and the transactions demand for cash. *Econometrica*, 77, 363-402.

Baumol, W. J., (1952). The transactions demand for cash: An inventory theoretic approach. *Quarterly Journal of Economics* 66, 545-556.

Benati L., R. E. Lucas Jr., J. P. Nicolini, and W. Weber, (2016). International evidence on long run money demand. *NBER Working Paper*, 22475.

Bordo, M. D., and E. U. Choudhri, (1982). Currency substitution and the demand for money: Some evidence for Canada. *Journal of Money, Credit and Banking* 14, 48-57.

Cagan, P., (1956). The monetary dynamics of hyperinflation. in *Studies in the Quantity Theory of Money*, M. Friedman (ed.), 25-117, Chicago: University of Chicago Press.

Chen, C-N., (1973). Diversified currency holdings and flexible exchange rates. *Quarterly Journal of Economics* 87, 96-111.

Chetty, V. K., (1969). On measuring the nearness of near-moneys. *American Economic Review* 59, 270-281.

Cuddington, J. T., (1983). Currency substitution, capital mobility and money demand. *Journal of International Money and Finance* 2, 111-133.

Dreger, C., H-E. Reimers, and B. Roffia, (2007). Long-run money demand in the new EU member states with exchange rate effects. *Eastern European Economics* 45, 75-94.

Elbourne, A., and J. de Haan, (2006). Financial structure and monetary policy transmission in transition countries. *Journal of Comparative Economics* 34, 1-23.

Fidrmuc, J., (2009). Money demand and disinflation in selected CEECs during the accession to the EU. *Applied Economics* 41, 1259-1267.

Filosa, R., (1995). Money demand stability and currency substitution in six European countries (1980: 1992). *BIS Working Paper*, 30.

Guidotti, P. E., (1989). Exchange rate determination, interest rates, and an integrative approach to the demand for money. *Journal of International Money and Finance* 8, 29-45.

Hansen, B. E., (1992). Tests for parameter instability in regressions with I(1) processes. *Journal of Business and Economic Statistics.* 10, 321-335.

Hsieh, W-J., and Y. Hsing, (2009). Tests of currency substitution, capital mobility and nonlinearity of Hungary's money demand function. *Applied Economics Letters* 16, 959-964.

Hueng, C. J., (1998). The demand for money in an open economy: Some evidence for Canada. *North American Journal of Economics and Finance* 9, 15-31.

Hueng, C. J., (2000). The impact of foreign variables on domestic money demand: Evidence from the United Kingdom. *Journal of Economics and Finance* 24, 97-109.

Ireland, P. N., (2009). On the welfare cost of inflation and the recent behavior of money demand. *American Economic Review* 99, 1040-1052.

Johansen, S., (1988). Statistical analysis of cointegrated vectors. *Journal of Economic Dynamics and Control* 12, 231-254.

Leventakis, J. A., (1993). Modelling money demand in open economies over the modern floating rate period. *Applied Economics* 25, 1005-1012.

Lucas, R., (1976). Econometric policy evaluation: A critique. in *The Phillips Curve and Labor Markets*, K. Brunner and A. Meltzer (eds.), 19-46, Carnegie-Rochester Conference Series on Public Policy 1. New York: American Elsevier.

Lucas, R., (1982). Interest rates and currency prices in a two country world. *Journal of Monetary Economics* 10, 335-359.

Lucas, R. E., Jr., (2000). Inflation and welfare. *Econometrica* 68, 247-274.

Lucas, R. E., Jr., and J. P. Nicolini, (2015). On the stability of money demand. *Journal of Monetary Economics* 73, 48-65.

Meltzer, A. H., (1963). The demand for money: The evidence from the time series. *Journal of Political Economy* 71, 219-246.

McKinnon, R. I., (1982). Currency substitution and instability in the world dollar standard. *American Economic Review* 72, 320-333.

Miles, M., (1978). Currency substitution, flexible exchange rates and monetary independence. *American Economic Review* 68, 428-436.

Miller, S. M., L. F. Martins, and R. Gupta, (forthcoming), A time-varying approach of the US welfare cost of inflation, *Macroeconomic Dynamics.*

Phillips, P., (1988). Reflections on econometric methodology. *Economic Record* 64, 344-359.

Phillips, P., (1991). Optimal inference in cointegrated systems. *Econometrica* 59, 283-306.

Phillips, P., and B. Hansen, (1990). Statistical inference in instrumental variables regression with I(1) processes. *Review of Economic Studies* 57, 99-125.

Saikkonen, P., (1991), Asymptotically efficient estimation of cointegrating regressions. *Econometric Theory* 7, 1-21.

Stock, J. H., and M. W. Watson, (1993). A simple estimator of cointegrating vectors in higher order integrated systems. *Econometrica* 61, 783-820.

Stockman, A., (1980). A theory of exchange rate determination. *Journal of Political Economy* 88, 673-698.

Tobin, J., (1956). The interest elasticity of the transactions demand for cash. *Review of Economics and Statistics* 38, 241-247.

Table 1. Unit root tests

| Variables | Tests | Czech Rep. | Hungary | Poland | Romania |
|---|---|---|---|---|---|
| $\ln\left(\dfrac{M_t}{S_t M_t^*}\right)$ | ADF | -1.788*** | -1.460*** | -0.808*** | -0.954*** |
| | PP | -1.674*** | -1.593*** | -0.769*** | -0.960*** |
| $\ln oc_{t+1} - \ln oc_{t+1}^*$ | ADF | -2.669** | -1.785*** | -1.371*** | -2.020*** |
| | PP | -2.661** | -2.350*** | -1.573*** | -2.058*** |
| $\ln\left(\dfrac{M2_t}{P_t}\right)$ | ADF | -0.012*** | -1.189*** | 1.065*** | -1.795*** |
| | PP | 0.048*** | -1.294*** | 0.752*** | -0.602*** |
| $\ln\left(\dfrac{M1_t}{P_t}\right)$ | ADF | -1.370*** | -1.365*** | -0.408*** | -1.650*** |
| | PP | -1.453*** | -1.162*** | -0.048*** | -0.685*** |
| $\ln oc_{t+1}$ | ADF | -1.181*** | -0.367*** | -0.852*** | -0.552*** |
| | PP | -1.269*** | -0.302*** | -1.016*** | -0.191*** |
| $\ln\left(\dfrac{XC_t}{P_t}\right)$ | ADF | -2.505*** | -2.202*** | -0.078*** | -1.070*** |
| | PP | -2.281*** | -2.573*** | -0.414*** | -1.470*** |
| $\ln\left(\dfrac{XIP_t}{P_t}\right)$ | ADF | -1.604*** | -1.396*** | -0.463*** | -7.620 |
| | PP | -3.114* | -2.023*** | -1.605*** | -8.420 |

*Notes: (i) the null hypothesis is the presence of unit root and \*, \*\*, \*\*\* means a p-value for the t-statistic >1%, >5% and >10% respectively. (ii) $\ln\left(\dfrac{M1_t}{P_t}\right)$ and $\ln\left(\dfrac{M2_t}{P_t}\right)$ are the two forms of $\ln\left(\dfrac{M_t}{P_t}\right)$ used in equation (20), considering the monetary aggregate M2 and M1, while $\ln\left(\dfrac{XC_t}{P_t}\right)$ and $\ln\left(\dfrac{XIP_t}{P_t}\right)$ are the two forms of $\ln\left(\dfrac{X_t}{P_t}\right)$ used in equation (20), considering the household consumption expenditure (C) and respectively the industrial production index (IP).*

Table 2. Cointegration test and estimations for equation (15)

| $\ln\left(\dfrac{M_t}{S_t M_t^*}\right)$ | Czech Rep. | | Hungary | | Poland | | Romania | |
|---|---|---|---|---|---|---|---|---|
| | DOLS | FMOLS | DOLS | FMOLS | DOLS | FMOLS | DOLS | FMOLS |
| $\kappa_0$ | 2.20*** | 2.17*** | 1.56*** | 1.43*** | 1.56*** | 1.69*** | 1.21*** | 0.93*** |
| $\kappa_1$ | 0.48*** | 0.38** | 0.27*** | 0.15 | -0.37** | -0.24 | 0.52*** | 0.32*** |
| $R^2$ | 0.30 | 0.16 | 0.25 | 0.06 | 0.29 | 0.08 | 0.74 | 0.48 |
| $L_c$ statistic | 0.00 (>0.2) | 1.28 (0.00) | 0.00 (>0.2) | 0.35 (0.10) | 0.00 (>0.2) | 1.83 (0.00) | 0.00 (>0.2) | 0.66 (0.01) |
| Wald t-statistic $\kappa_1 = 1$ | 2.83 (0.00) | 3.92 (0.00) | 6.00 (0.00) | 8.72 (0.00) | 7.58 (0.00) | 7.36 (0.00) | 6.18 (0.00) | 10.5 (0.00) |

*Notes: (i) \*\*\*, \*\*, \* means significance at 1%, 5% et 10% significance level; (ii) p-value in brackets; (iii) $\kappa_0$ is the intercept of equation (15); $\kappa_1$ is the coefficient of $\left(\ln oc_{t+1} - \ln oc_{t+1}^*\right)$, with a negative sign.*

Table 3. Cointegration test and estimations for equation (20)

| $\ln\left(\dfrac{M2_t}{P_t}\right)$ | Czech Rep. | | Hungary | | Poland | | Romania | |
|---|---|---|---|---|---|---|---|---|
| **Consumption** | DOLS | FMOLS | DOLS | FMOLS | DOLS | FMOLS | DOLS | FMOLS |
| $\omega_0$ | -7.68*** | -2.30 | 1.92* | -0.02 | -7.92*** | -5.65*** | -1.44*** | -3.80*** |
| $\omega_1$ | 0.07* | 0.08** | 0.23*** | 0.19*** | -0.21*** | -0.04 | 0.20*** | 0.13*** |
| $\omega_2$ | 0.90*** | 0.23*** | 0.02 | 0.10*** | 0.01 | 0.05** | 0.03* | 0.17*** |
| $\omega_3$ | 2.45*** | 1.61*** | 0.89*** | 1.11*** | 2.72*** | 2.23*** | 1.39*** | 1.83*** |
| $R^2$ | 0.98 | 0.88 | 0.98 | 0.80 | 0.99 | 0.98 | 0.99 | 0.98 |
| $L_c$ statistic | 0.00 (>0.2) | 1.01 (0.02) | 0.00 (>0.2) | 0.68 (0.09) | 0.00 (>0.2) | 2.02 (0.00) | 0.00 (>0.2) | 0.93 (0.03) |
| Wald t-statistic $\omega_1 = 1$ | 27.3 (0.00) | 21.5 (0.00) | 34.2 (0.00) | 29.0 (0.00) | 24.0 (0.00) | 28.6 (0.00) | 46.9 (0.00) | 32.3 (0.00) |
| Wald t-statistic $\omega_3 = 1$ | 5.36 (0.00) | 2.66 (0.00) | 0.95 (0.34) | 0.76 (0.44) | 11.2 (0.00) | 10.1 (0.00) | 5.64 (0.00) | 10.4 (0.00) |
| **Industrial production** | DOLS | FMOLS | DOLS | FMOLS | DOLS | FMOLS | DOLS | FMOLS |
| $\omega_0$ | 9.38*** | 8.65*** | 6.60*** | 9.60*** | 8.75*** | 7.56*** | 2.98*** | 4.48*** |
| $\omega_1$ | 0.08 | 0.21*** | 0.79*** | 0.36*** | -0.02 | 0.19*** | 0.97*** | 0.53*** |
| $\omega_2$ | 5.93*** | 0.41*** | 0.73*** | -0.19 | 0.26*** | 0.24*** | -0.35*** | 0.10 |
| $\omega_3$ | 2.14*** | 1.45*** | 3.33*** | 0.68 | 2.27*** | 1.68*** | -1.14*** | -0.46* |
| $R^2$ | 0.95 | 0.78 | 0.90 | 0.45 | 0.99 | 0.91 | 0.98 | 0.83 |
| $L_c$ statistic | 0.00 (>0.2) | 0.49 (>0.2) | 0.00 (>0.2) | 1.66 (0.00) | 0.00 (>0.2) | 0.50 (>0.2) | 0.00 (>0.2) | 1.28 (0.00) |
| Wald t-statistic $\omega_3 = 1$ | 2.05 (0.04) | 1.70 (0.09) | 3.35 (0.00) | 0.60 (0.58) | 3.95 (0.00) | 3.48 (0.00) | 5.87 (0.00) | 5.46 (0.00) |

*Notes: (i) \*\*\*, \*\*, \* means significance at 1%, 5% et 10% significance level; (ii) p-value in brackets; (iii) $\omega_0$ is the intercept of equation (20); (iv) $\omega_1$ is the coefficient of $\ln oc_{t+1}$ with a negative sign, $\omega_2$ is the coefficient of $\left(\ln oc_{t+1} - \ln oc^*_{t+1}\right)$ and $\omega_3$ is the coefficients of $\ln\left(\dfrac{XC_t}{P_t}\right)$ or $\ln\left(\dfrac{XIP_t}{P_t}\right)$; (v) M2 aggregate is considered.*

Table 4. Cointegration test and estimations for equation (20) – robustness check based on M1

| $\ln\left(\dfrac{M1_t}{P_t}\right)$ | Czech Rep. | | Hungary | | Poland | | Romania | |
|---|---|---|---|---|---|---|---|---|
| **Consumption** | DOLS | FMOLS | DOLS | FMOLS | DOLS | FMOLS | DOLS | FMOLS |
| $\omega_0$ | -12.0*** | -11.2*** | 2.69*** | -0.60 | -11.3*** | -3.80*** | -9.50*** | -11.8*** |
| $\omega_1$ | 0.31*** | 0.34*** | 0.50*** | 0.36*** | 0.09** | 0.13*** | 0.06 | -0.00 |
| $\omega_2$ | 0.43* | 0.15*** | -0.08*** | 0.04 | -0.10*** | 0.17*** | -0.00 | 0.12 |
| $\omega_3$ | 2.68*** | 2.55*** | 0.61*** | 1.01 | 2.92*** | 1.83*** | 2.82*** | 3.26*** |
| $R^2$ | 0.99 | 0.96 | 0.98 | 0.87 | 0.99 | 0.98 | 0.99 | 0.96 |
| $L_c$ statistic | 0.00 (>0.2) | 0.95 (0.02) | 0.00 (>0.2) | 0.55 (0.17) | 0.00 (>0.2) | 0.93 (0.02) | 0.00 (>0.2) | 0.55 (0.17) |
| Wald t-statistic $\omega_3 = 1$ | 4.75 (0.00) | 7.05 (0.00) | 4.14 (0.00) | 0.07 (0.93) | 12.3 (0.00) | 10.4 (0.00) | 3.40 (0.00) | 9.49 (0.00) |
| **Industrial production** | DOLS | FMOLS | DOLS | FMOLS | DOLS | FMOLS | DOLS | FMOLS |
| $\omega_0$ | 6.95*** | 6.18*** | 5.23*** | 8.00*** | 6.86*** | 4.48*** | -0.62 | 2.43** |
| $\omega_1$ | 0.40*** | 0.55*** | 0.96*** | 0.53*** | 0.22*** | 0.53*** | 1.62*** | 0.77*** |
| $\omega_2$ | 0.77*** | 0.42*** | 0.53*** | 0.17 | 0.17*** | 0.10 | -0.79*** | -0.01 |
| $\omega_3$ | 3.38*** | 2.31*** | 3.04*** | 0.85* | 2.73*** | -0.46* | -1.66** | -0.57 |
| $R^2$ | 0.99 | 0.89 | 0.96 | 0.70 | 0.99 | 0.83 | 0.97 | 0.75 |
| $L_c$ statistic | 0.00 (>0.2) | 0.58 (0.15) | 0.00 (>0.2) | 1.61 (0.00) | 0.00 (>0.2) | 1.28 (0.00) | 0.00 (>0.2) | 0.96 (0.02) |
| Wald t-statistic $\omega_3 = 1$ | 5.27 (0.00) | 4.36 (0.00) | 4.63 (0.00) | 0.31 (0.75) | 8.20 (0.00) | 5.46 (0.00) | 3.63 (0.00) | 3.06 (0.00) |

*Notes: (i) \*\*\*, \*\*, \* means significance at 1%, 5% et 10% significance level; (ii) p-value in brackets; (iii) $\omega_0$ is the intercept of equation (20); (iv) $\omega_1$ is the coefficient of $\ln oc_{t+1}$ with a negative sign, $\omega_2$ is the coefficient of $\left(\ln oc_{t+1} - \ln oc_{t+1}^*\right)$ and $\omega_3$ is the coefficients of $\ln\left(\dfrac{XC_t}{P_t}\right)$ or $\ln\left(\dfrac{XIP_t}{P_t}\right)$.*

**Appendix - Data description**

| Variables | Database | Explanations |
|---|---|---|
| M2 | IFS (IMF) | Monetary aggregate M2 in millions of national currency. As data are not available for all the countries over the entire time-span 1999:M1-2015M11, the series was completed as follows: OECD statistics for the Czech Republic (1999-2001) and for Hungary (1999-2004); national bank statistics for Romania (1999-2001). |
| M1 | IFS (IMF) | Monetary aggregate M1 in millions of national currency. As data are not available for all the countries over the entire time-span 1999:M1-2015M11, the series was completed as follows: OECD statistics for the Czech Republic (1999-2001); national bank statistics for Romania (1999-2001). |
| Interest rate | IFS (IMF) | The money market rate from the International Financial Statistics. For Hungary and the euro area, Eurostat data (day-to-day). |
| Domestic deposits to foreign deposits ratio | National Banks | For the Czech Republic, banking clients' deposits in foreign currency to total deposits. For Hungary, Poland and Romania, aggregated balance sheet data of credit institutions. |
| Prices | IFS (IMF) | Consumer Price Index (2010=100). |
| Consumption | IFS (IMF) | Household consumption expenditure (quarterly data transformed in monthly data using a cubic spline function). |
| Output | IFS (IMF) | Industrial Production index (2010=100). |